\documentclass[12pt]{article}
\usepackage[onehalfspacing]{setspace}
\usepackage{tipa}
\usepackage{graphicx}
\usepackage{caption}
\usepackage{longtable}
\usepackage{tabularx}
\usepackage{mathptmx}
\usepackage{mathtools}
\usepackage{multirow}
\usepackage{array}
\usepackage{graphicx}
\usepackage{subcaption}
\usepackage[numbers]{natbib} 
\usepackage{amsmath}

\usepackage{amssymb}
\usepackage{amsfonts}
\usepackage{amsthm}
\usepackage{mathtools}
\usepackage{setspace}
\usepackage{stmaryrd}
\usepackage{epstopdf}  
\usepackage{grffile} 
\usepackage[margin=1in]{geometry}
\usepackage{amsfonts}
\usepackage{booktabs}
\usepackage{url}
\usepackage{hyperref}
\usepackage{float}
\usepackage{tikz}
\usepackage{fancyhdr}
\usepackage{authblk}
\usepackage [utf8]{inputenc}

\title{\vspace{-1.2in} Individualized Treatment Effects in Advanced Prostate Cancer: A Causal-Survival Modeling Approach to Risk-Guided Therapy}
\author[1]{Joshua Korley}
\affil[1]{Arnold School of Public Health, University of South Carolina, Columbia, SC, USA}
\date{}

\begin{document}

\maketitle
\begin{abstract}
We conducted a proof-of-concept evaluation of individualized treatment effect (ITE) estimation using survival data from a randomized trial of 475 men with advanced prostate cancer treated with high- versus low-dose diethylstilbestrol (DES). A Weibull accelerated failure time (AFT) model with interaction terms for treatment-by-age and treatment-by-log tumor size was used to capture subgroup-specific treatment effects. The estimated main effect of high-dose DES indicated a time ratio of 0.582 (95\% CI: [0.306, 1.110]), reflecting reduced survival at the reference levels of age and tumor size. However, interaction-adjusted ITEs revealed marked effect modification: younger patients (e.g., age 50 years) had over fourfold expected survival gains (time ratio 4.09), whereas older patients (e.g., age 80 years) experienced reduced benefit (time ratio 0.71). Similarly, patients with larger tumors (log size $\sim$4.25, $\sim$70 cm\textsuperscript{2}) derived a stronger benefit (time ratio 1.89) than those with smaller tumors. To evaluate the reliability of these individualized estimates, both the delta method and bootstrap resampling were applied for uncertainty quantification, producing closely aligned intervals across the risk spectrum. This analysis illustrates how parametric survival models with clinically motivated interactions and robust inference procedures can yield interpretable patient-level treatment effect estimates, even in moderately sized oncology trials.
\\

 \textbf{Keywords:} \tiny
 Individualized Treatment Effect, Survival Analysis, Delta Method, Bootstrap, Causal Inference, Prostate Cancer, Weibull AFT Model
\end{abstract}

\section{Introduction}

Prostate cancer is one of the most frequently diagnosed malignancies in men and a major contributor to cancer-related mortality, particularly among older adults \citep{crawford2003epidemiology, jemal2010cancer}. In advanced disease stages, hormonal therapies remain the cornerstone of treatment, with diethylstilbestrol (DES), a synthetic estrogen, historically used to suppress androgen production. Although high-dose DES has demonstrated antitumor activity, concerns regarding cardiovascular toxicity have led to greater reliance on low-dose regimens or alternative hormonal agents \citep{keating2006diabetes, kumar2005adverse}. The trade-off between efficacy and toxicity becomes especially complex in heterogeneous patient populations with differing tumor burdens and comorbidities \citep{heinzer2009prostate, cooperberg2009risk}.

Despite decades of clinical research, treatment decisions for advanced prostate cancer often rely on population-level estimates of efficacy, potentially obscuring important inter-individual differences in treatment response. In practice, the application of uniform therapy across biologically diverse subgroups may lead to suboptimal outcomes, especially in older or comorbid patients who face competing health risks \citep{cooperberg2004changing, alibhai2003continued}. Precision oncology seeks to address this challenge by tailoring treatment to individual characteristics; however, many survival analyses remain anchored in average-effect models that do not fully exploit the potential of individualized decision-making.

Most historical analyses in prostate cancer trials have utilized Cox proportional hazards (PH) models to evaluate treatment efficacy \citep{byar1973veterans, beksisa2020survival}. Although widely used, the Cox model assumes proportional hazards and yields hazard ratios that are difficult to interpret when interaction or time-varying effects are present. These limitations become more problematic when the goal is to estimate patient-specific treatment effects rather than marginal or average outcomes \citep{hernan2010hazards, aalen2008survival}.

Accelerated failure time (AFT) models offer a more flexible alternative that directly models survival time and provides interpretable time ratios than the Cox model. The Weibull AFT model is capable of accommodating both proportional and non-proportional hazards, making it suitable for individualized treatment effect (ITE) estimation in heterogeneous populations. Moreover, it allows the formal incorporation of treatment-covariate interaction terms, facilitating the estimation of how treatment efficacy varies across patient characteristics.

In this study, we revisited data from a randomized trial of high- versus low-dose DES in advanced prostate cancer to estimate both the average and individualized treatment effects using a Weibull AFT model. Our modeling strategy explicitly incorporates treatment-by-covariate interaction terms for baseline age and log-transformed tumor size, which are established prognostic and predictive markers of prostate cancer \citep{berney2016prognostic, tangen2012longterm}. Using this framework, we define patient-specific log time ratio estimands \(\hat{\Delta}_i\) under a potential outcomes formulation \citep{rubin1974estimating}, enabling individualized survival contrasts between treatment arms.

In addition to estimating \(\hat{\Delta}_i\), we adopted two complementary approaches to quantify the uncertainty of these patient-level effects: (i) the delta method, which yields closed-form confidence intervals based on the asymptotic distribution of model estimates, and (ii) nonparametric bootstrap resampling, which allows for empirical estimation of variability. Few studies have applied both methods in tandem for ITE inference in survival analysis, despite their distinct inferential properties and interpretability implications. Our methodological framework aligns with recent simulation-based studies that evaluate risk-based estimands in randomized trials \citep{rek2023bmc}.

This study contributes a transparent, interpretable, and reproducible pipeline for estimating individualized treatment effects in oncology. By moving beyond average-effect modeling and explicitly quantifying uncertainty in patient-level predictions, we aim to support more informed clinical decision-making in prostate cancer, particularly in the context of risk-adaptive treatment selection.

\section{Methods and Materials}

\subsection{Study Design and Data Description}
This study analyzed data from a randomized clinical trial that investigated survival outcomes in 475 men diagnosed with advanced prostate cancer. Participants were randomly assigned to receive either a high (5 mg/day) or low (1 mg/day) dose of diethylstilbestrol (DES). The primary endpoint was overall survival, defined as the time from enrollment to death, and was measured in months. Censoring was applied to individuals who were alive at the last follow-up visit. Randomization and complete baseline measurement collection permit causal comparisons of treatment effects under minimal assumptions.

\subsection{Covariates and Preprocessing}

The original dataset comprised 15 baseline covariates. For modeling purposes, we retained a parsimonious subset based on clinical relevance, distributional diagnostics, and the results of univariate Cox regression analyses (Table~\ref{tab:Table 2}). The final model included age (years), weight (kg; standardized), serum hemoglobin (g/100 ml), size of the primary tumor (log-transformed cm\textsuperscript{2}), a numeric index combining tumor stage and histologic grade, serum prostatic acid phosphatase (King-Armstrong units), history of cardiovascular disease (binary), electrocardiogram code (ordinal: 0–6), and cancer stage (Stage 3 vs. Stage 4).

Several variables were excluded to improve the model’s stability and interpretability. Performance ratings and bone metastases, while clinically meaningful, exhibited extreme sparsity and quasi-complete separation in preliminary models, resulting in inflated standard errors and convergence issues. Systolic and diastolic blood pressure were excluded due to non-significant associations with survival in univariate analyses ($p = 0.669$ and $p = 0.43$, respectively), and to avoid redundancy in vascular risk representation, which was already captured via the cardiovascular disease history. Serum acid phosphatase was similarly excluded due to the lack of a univariate association ($p = 0.581$). Covariate selection was conducted prior to model fitting to minimize overfitting and preserve the interpretability of the interaction terms.

Treatment was coded as a binary indicator: 1 = high-dose DES and 0 = low-dose DES treatment.

\subsection{Modeling Strategy and Causal Framework}

We adopted the potential outcomes framework to define individualized treatment effects (ITEs) under the Weibull accelerated failure time (AFT) model. Let $Z_i \in \{0, 1\}$ indicate randomized treatment assignment for subject $i$, where $Z_i = 1$ denotes high-dose diethylstilbestrol (DES), and $Z_i = 0$ denotes low-dose DES. Let $T_i(1)$ and $T_i(0)$ represent the potential survival times under high- and low-dose treatment, respectively. Under the assumptions of consistency, stable unit treatment value assumption (SUTVA), and conditional exchangeability (guaranteed by randomization), the observed survival time is $T_i = T_i(Z_i)$ and

\begin{equation}
\{T_i(1), T_i(0)\} \perp Z_i \mid X_i
\end{equation}

where $X_i$ is the vector of baseline covariates. The causal estimand of interest is the patient-specific log time ratio.
\begin{equation}
\Delta_i = \log T_i(1) - \log T_i(0),
\end{equation}
which compared the log survival time under high-dose versus low-dose DES. A positive value of $\Delta_i$ indicates a longer expected survival under high-dose treatment.

To model the log survival time, we employed a parametric AFT model assuming a Weibull distribution:
\begin{equation}
\log T_i = \mu + \beta_Z Z_i + \boldsymbol{\beta}_X^\top X_i + \sum_{j \in \mathcal{M}} \gamma_j Z_i X_{ij} + \varepsilon_i,
\end{equation}
where $\mu$ is the intercept, $\beta_Z$ is the marginal treatment effect, $\boldsymbol{\beta}_X$ are coefficients for baseline covariates, $\gamma_j$ are interaction terms indexed by the subset $\mathcal{M}$ of covariates modifying the treatment effect, and $\varepsilon_i$ follows an extreme value distribution (type I), corresponding to the Weibull survival model.

Three models were specified to evaluate the treatment effects and identify relevant modifiers.
\begin{itemize}
    \item[1.]  Model 1 included the main effects of all covariates and the treatment indicator and all treatment-by-covariate interaction terms. 
    \item[2.] Model 2A retained treatment, all covariate main effects, and interaction terms only for two pre-specified clinical modifiers: age and log-transformed tumor size. This model reflects a clinically constrained formulation based on exploratory diagnostics and prior evidence of effect heterogeneity.
    \item[3.] Model 2B was the main effect for all covariates, the treatment indicator, and all treatment-by-covariate interaction terms. This full. Variables exhibiting multicollinearity, quasi-complete separation, or data sparsity were excluded to ensure the model stability.
\end{itemize}

Model 2A was selected as the primary analysis model based on its goodness-of-fit and parsimony. All individualized treatment effect estimations were derived from this model.

Under Model 2A, the ITE for subject $i$ is given by
\begin{equation}
\hat{\Delta}_i = \hat{\beta}_Z + \sum_{j \in \mathcal{M}} \hat{\gamma}_j X_{ij},
\end{equation}
The corresponding time ratio is $\exp(\hat{\Delta}_i)$, representing the multiplicative change in expected survival time under high-dose versus low-dose DES. Values of $\exp(\hat{\Delta}_i) > 1$ indicate an individual-level benefit from high-dose therapy.

To quantify the uncertainty in $\hat{\Delta}_i$, we implemented two complementary inference strategies:
\begin{enumerate}
    \item The delta method (which is parametric), which propagates variance from the model’s asymptotic covariance matrix to derive analytic confidence intervals.
    \item A nonparametric bootstrap procedure, resampling patients with replacement and refitting the AFT model across 1,000 iterations to construct empirical confidence intervals.
\end{enumerate}

This dual approach to inference ensures a robust characterization of patient-specific uncertainty and complements the causal interpretation of ITEs under the AFT framework.

\subsection{Delta Method and Bootstrap}
To quantify the uncertainty in $\hat{\Delta}_i$, we used both the delta method and nonparametric bootstrap method.

\subsubsection{Delta Method}
Let $\boldsymbol{\theta} = (\mu, \beta_Z, \boldsymbol{\beta}_X, \boldsymbol{\gamma})$ denote the full vector of model parameters, and let $\hat{\Sigma}$ be the estimated covariance matrix of $\hat{\boldsymbol{\theta}}$. For a given patient $i$, let $\boldsymbol{v}_i = \partial \Delta_i / \partial \boldsymbol{\theta}$ represent the gradient of $\Delta_i$ with respect to $\boldsymbol{\theta}$, which selects only the components associated with the treatment effect and its interaction(s).

When estimating ITEs by age, we used a model that included the treatment main effect and the interaction with age: $\Delta_i = \beta_{\text{rx}} + \gamma_{\text{rx:age}} \cdot \text{age}_i$. The gradient vector was $\boldsymbol{v}_i = [0, 1, 0, \ldots, 0, \text{age}_i, 0, \ldots, 0]$, with non-zero entries corresponding to $\beta_{\text{rx}}$ and $\gamma_{\text{rx:age}}$.

Similarly, for ITEs by log tumor size, the model used $\Delta_i = \beta_{\text{rx}} + \gamma_{\text{rx:logsz}} \cdot \log(\text{tumor size}_i)$ and $\boldsymbol{v}_i = [0, 1, 0, \ldots, 0, \log(\text{tumor size}_i), 0, \ldots, 0]$.

The variance was estimated as follows:
\begin{equation}
\text{Var}(\hat{\Delta}_i) \approx \boldsymbol{v}_i^\top \hat{\Sigma} \boldsymbol{v}_i
\end{equation}
We then computed 95\% confidence intervals as follows:
\begin{equation}
\hat{\Delta}_i \pm z_{0.975} \cdot \sqrt{\text{Var}(\hat{\Delta}_i)}
\end{equation}
with $z_{0.975}$ being the 97.5th percentile of standard normal distribution. These were exponentiated to obtain the confidence intervals for $\exp(\hat{\Delta}_i)$. This methodology follows the approach described by \cite{hosmer2011applied}.

\subsubsection{Bootstrap Method}
To complement and validate the analytical delta method, we conducted a nonparametric bootstrap analysis as follows.
\begin{enumerate}
  \item Generate 1,000 bootstrap resamples by sampling patients with replacement.
  \item Fit the selected AFT model (Model 2A) to each resample.
  \item For each patient, compute $\hat{\Delta}_i^{(b)}$ under the fitted model.
  \item Construct 95\% percentile confidence intervals using the 2.5th and 97.5th percentiles of the bootstrap distribution $\{\hat{\Delta}_i^{(1)}, \ldots, \hat{\Delta}_i^{(1000)}\}$.
\end{enumerate}
This approach accounts for variability in both the data and model-fitting processes, providing robust inferences.

\subsection{Model Selection and Validation}

To identify an appropriate model for estimating individualized treatment effects, we first compared three parametric AFT distributions—log-normal, log-logistic, and Weibull—using the information criteria. The Weibull model yielded the lowest AIC (3260.22) and BIC (3314.34), favoring it over the log-normal and log-logistic alternatives (Table~\ref{tab:model_comp}). The Weibull distribution also accommodates monotonic hazard shapes and is clinically interpretable, making it suitable for modeling prostate cancer progression.
\begin{figure}[htbp]
  \centering
  \includegraphics[width=0.9\textwidth, height=7cm]{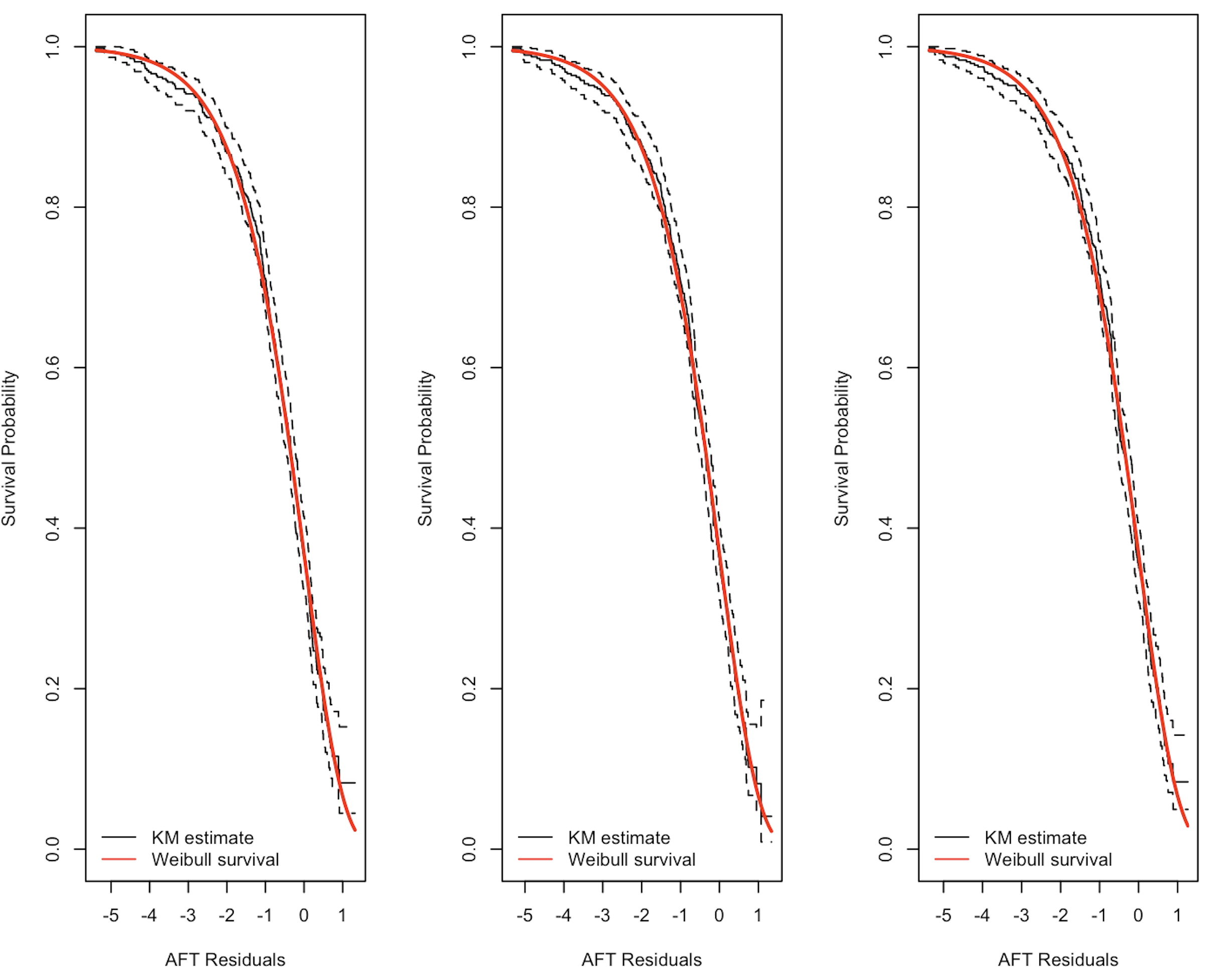}
  \caption{Weibull AFT residual diagnostics. Kaplan--Meier estimate (black) and model-implied Weibull survival (red) over AFT residuals for Model~1 (left), Model~2A (middle), and Model~2B (right). Closer agreement between curves indicates better fit.}
  \label{fig:aft-resid-weibull}
\end{figure}

We then evaluated the three model specifications under the Weibull AFT framework. Model 1 included treatment and all baseline covariates as main effects only, whereas Model 2B extended this by adding interaction terms between treatment and all covariates. Model 2A retained treatment and all covariate main effects but restricted treatment interactions to age and log-transformed tumor size based on prior evidence and diagnostics.

\begin{figure}[htbp]
  \centering
  \includegraphics[width=\textwidth, height= 8cm]{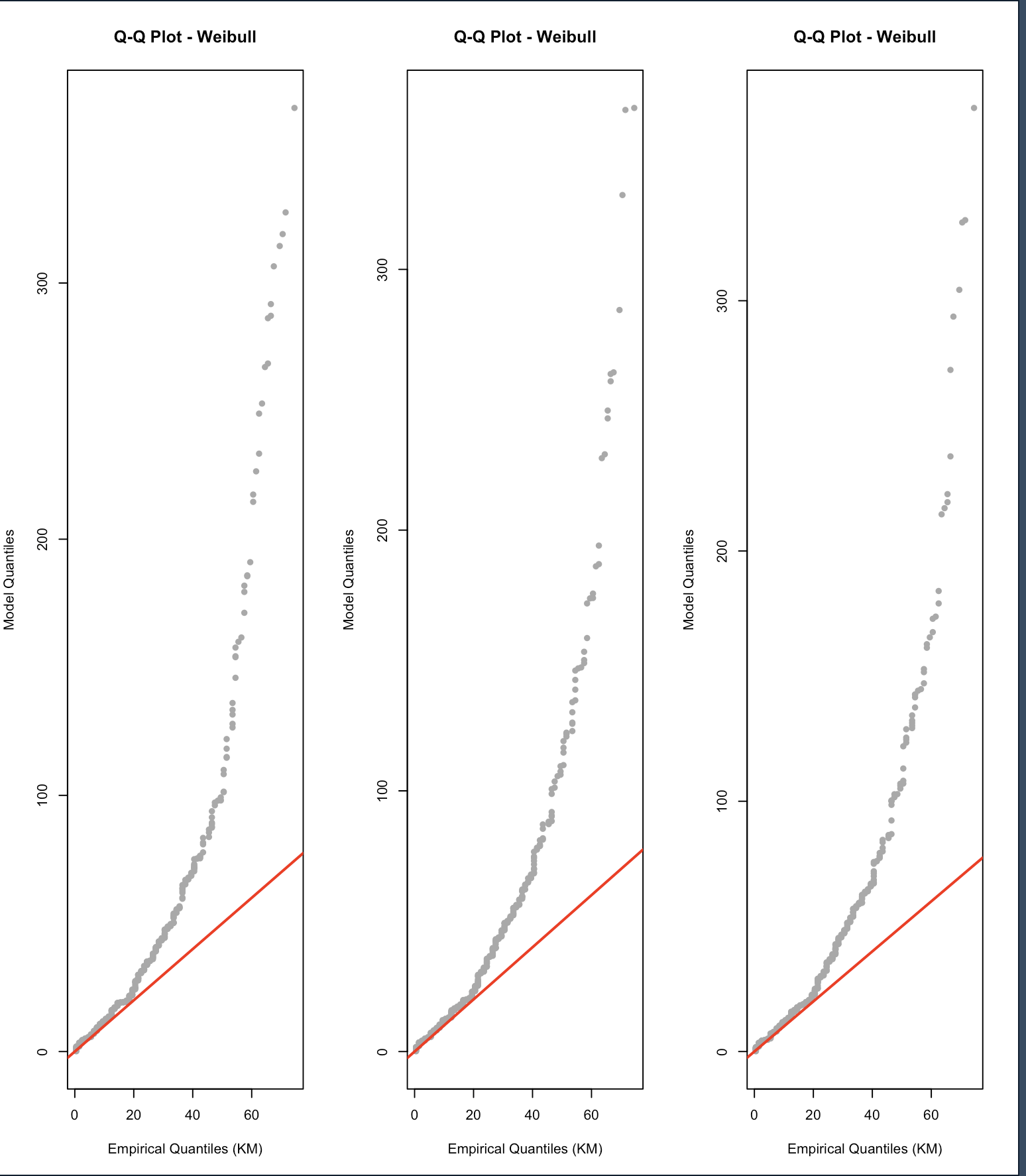}
  \caption{Weibull AFT Q--Q diagnostics. Model quantiles vs. empirical (Kaplan--Meier) quantiles with 45$^\circ$ reference line (red) for Model~1 (left), Model~2A (middle), and Model~2B (right). Closer alignment to the line indicates better distributional fit.}
  \label{fig:qq-weibull}
\end{figure}

Model 2A demonstrated a superior fit, with the lowest AIC and BIC (Table~\ref{tab:model_comp_poster}), outperforming the more saturated Model 2B and the simpler Model 1. Residual-based diagnostics confirmed this selection process. Kaplan–Meier overlays of the AFT residuals showed that Model 2A achieved the closest alignment between the empirical and fitted survival curves (Figure~\ref{fig:aft-resid-weibull}). The Q–Q plots comparing the model and empirical quantiles likewise showed the best conformity under Model 2A (Figure~\ref{fig:qq-weibull}).

These results established Model 2A as the most appropriate working model for estimating treatment effect heterogeneity while balancing clinical relevance, interpretability, and statistical fit.

\subsection{Analysis Workflow}

All analyses were performed using \texttt{R} version 4.4.1. Parametric survival models were estimated using the \texttt{flexsurv} package with model-based standard errors and distribution-specific diagnostics. Delta method variance estimates were computed using the \texttt{car} package, whereas nonparametric bootstrap inference was implemented using custom scripts. All visualizations, including survival overlays, quantile–quantile diagnostics, and individualized treatment effect plots, were generated using \texttt{ggplot2} and \texttt{survminer}.

\begin{figure}[htbp]
  \centering
  \includegraphics[width=\textwidth]{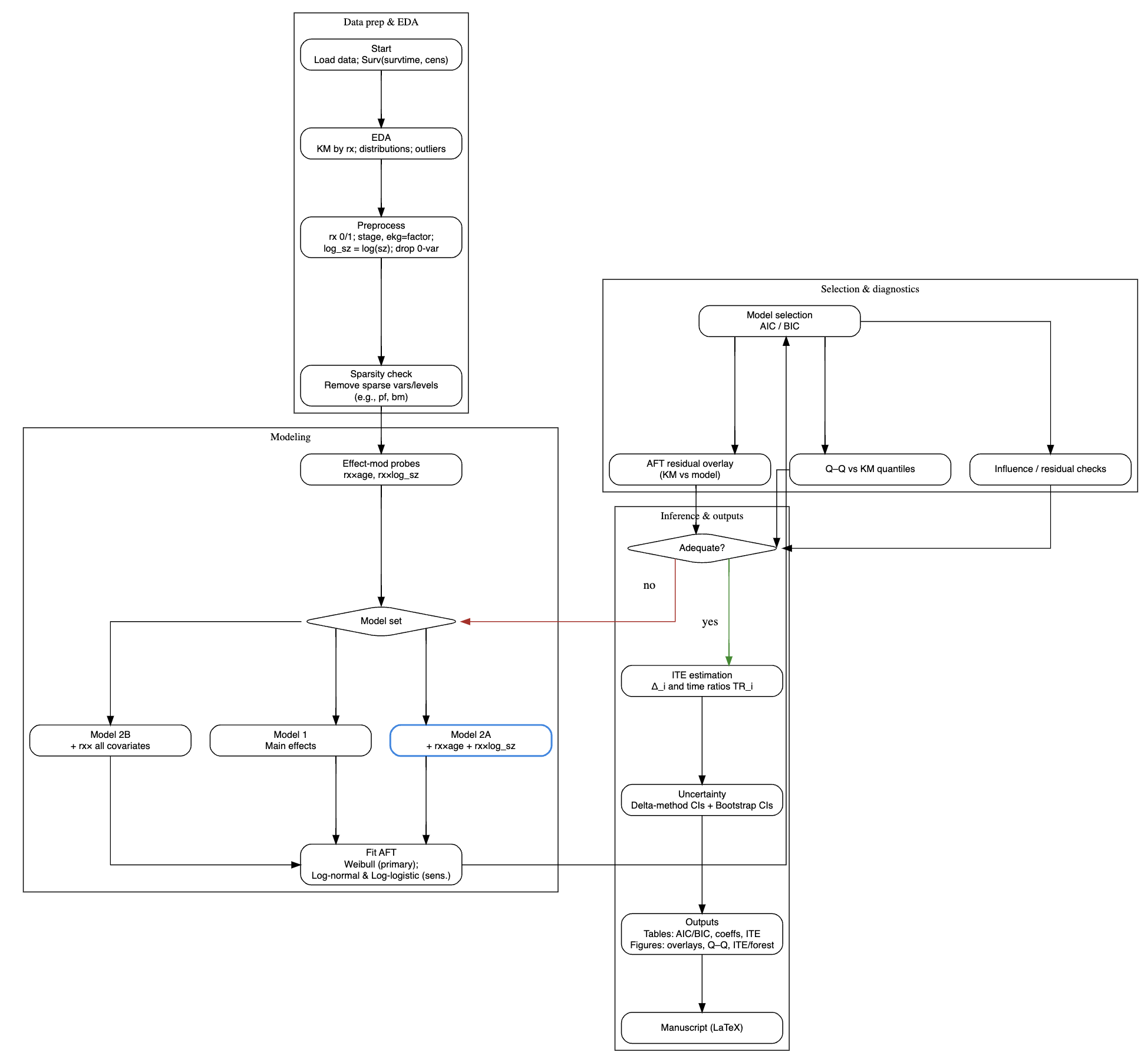}
  \caption{Comprehensive analysis workflow: data preprocessing, parametric modeling, diagnostic checks, and individualized treatment effect estimation.}
  \label{fig:analysis-flow}
\end{figure}

The complete modeling and inference pipeline is summarized in Figure~\ref{fig:analysis-flow}, encompassing pre-processing, model selection, residual diagnostics, and estimation of patient-specific treatment effects. All steps from sparsity checks and interaction probing to parametric model fitting, uncertainty quantification, and final table/figure outputswere reproducibly implemented within a structured \texttt{R} workflow.

\newpage
\section{Results}

\subsection{Descriptive Analysis and Kaplan--Meier Survival Comparison}

The dataset consisted of 475 men with advanced prostate cancer who were randomly assigned to receive low-dose (n = 237) or high-dose (n = 238) diethylstilbestrol (DES). The median age at enrollment was 68 years (range: 45–88 years), and 338 deaths (71.2\%) were observed during follow-up (Table~\ref{tab:treat_diag}). The mean tumor size at baseline was 14.29 cm\textsuperscript{2} (standard deviation = 9.2). Baseline characteristics, including tumor stage, performance rating, cardiovascular disease history, hemoglobin level, blood pressure, and electrocardiogram codes, exhibited no meaningful imbalances across the treatment groups, consistent with successful randomization.

\begin{figure}[h!]
\begin{center}
\includegraphics[width=18cm]{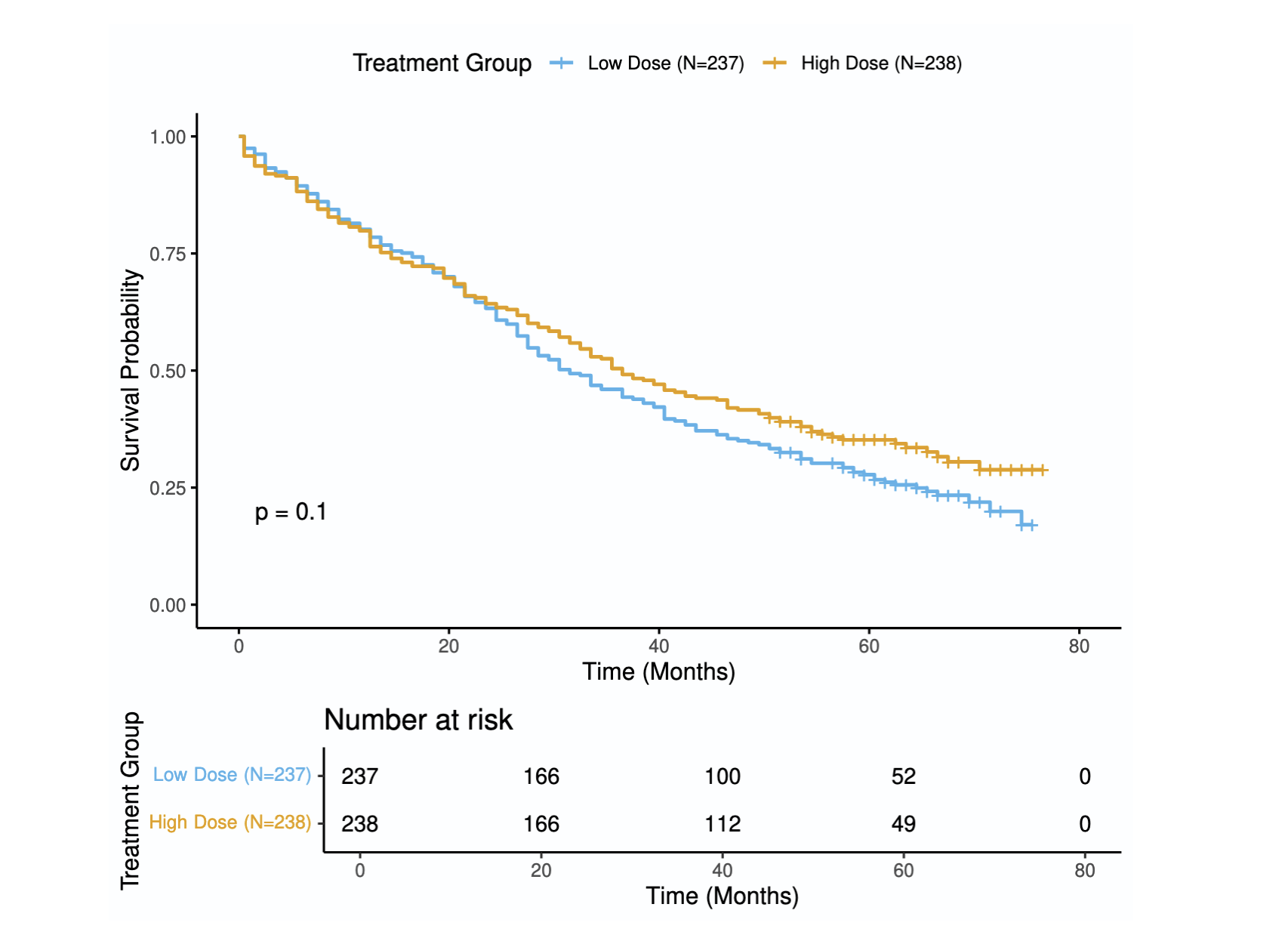}%
\end{center}
\caption{ Kaplan-Meier survival curves comparing low-dose vs. high-dose DES groups. The survival probability over time is stratified by treatment arm.}\label{fig: Figure_1}
\end{figure}

Initial survival comparisons were conducted using Kaplan--Meier curves stratified by the treatment arm (Figure~\ref{fig: Figure_1}). While the high-dose DES group displayed modestly elevated survival probabilities throughout the follow-up period, the difference was not statistically significant based on the log-rank test ($p = 0.10$). At 60 months, the estimated survival probabilities were approximately 0.35 and 0.28 in the high-and low-dose groups, respectively. These findings suggest a potential treatment-related difference in survival, which warrants further investigation through covariate-adjusted modeling.

\subsection{Covariate-Adjusted Average Treatment Effects}

\begin{figure}[h!]
    \centering
    \includegraphics[width=0.9\linewidth]{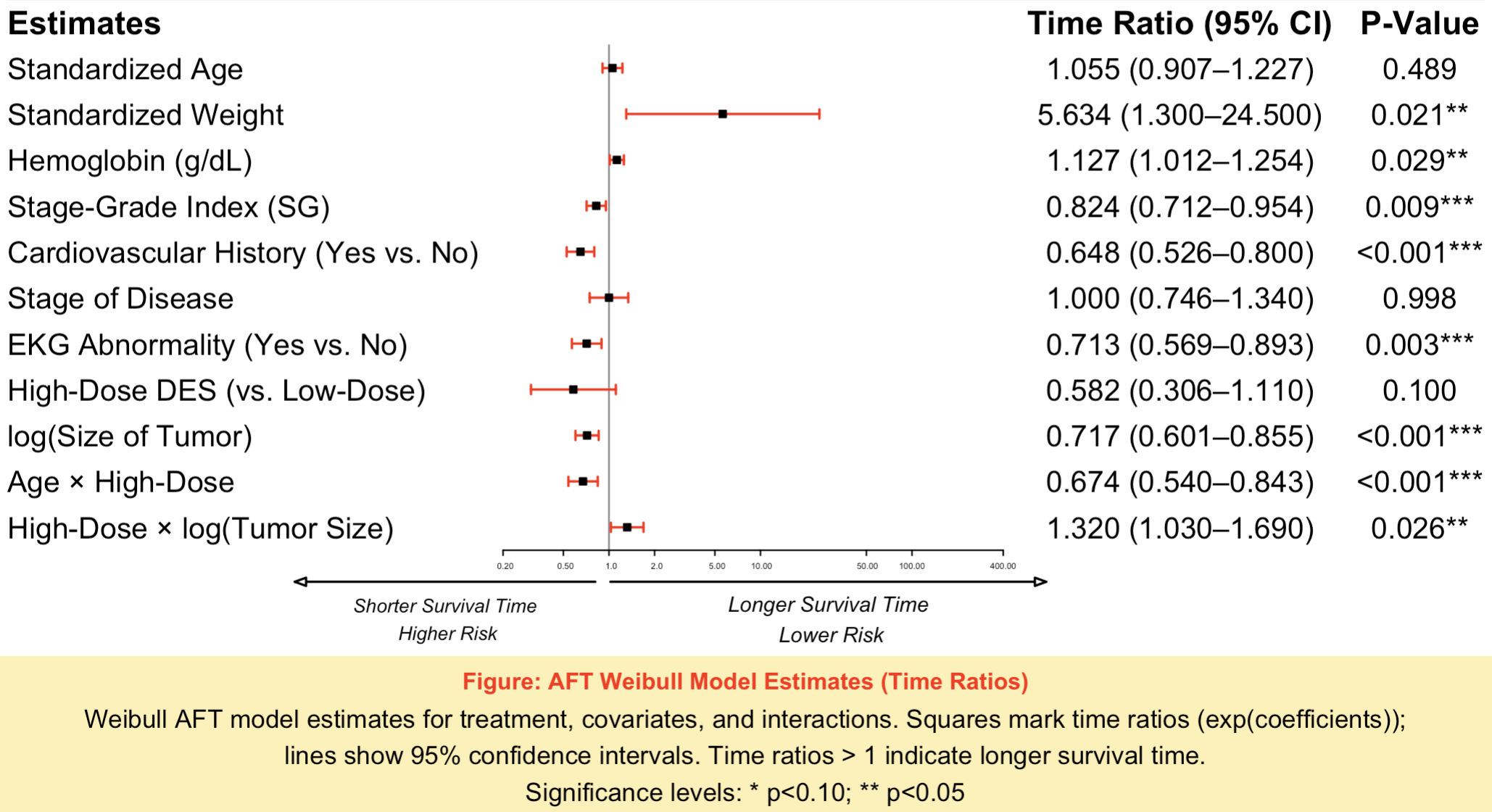}
    \caption{Forest plot of Model 2A showing time ratios and 95\% confidence intervals for treatment and key covariates.}
    \label{fig:ate}
\end{figure}

To quantify the treatment effect accounting for patient-level covariates, we fitted a Weibull Accelerated Failure Time (AFT) model (Model 2A), which included treatment, age, weight, hemoglobin, tumor stage, log tumor size, electrocardiogram code, and interaction terms for treatment-by-age and treatment-by-log tumor size. Parameter estimates from this model are summarized in Figure~\ref{fig:ate}.

The coefficient for high-dose DES yielded a time ratio of 0.582 (95\% CI: [0.306, 1.110]; $p = 0.100$), indicating a non-significant average treatment effect on survival rate. However, the interaction between treatment and age was positive and highly significant ($p < 0.001$), whereas the interaction between treatment and log tumor size was negative and statistically significant ($p = 0.026$). These terms reflect differential treatment effects that depend on patient age and tumor burden, motivating the development of an individualized treatment effect (ITE) estimation framework.

\subsection{Effect Modification and Heterogeneous Treatment Effects}

The estimated coefficients for the interaction terms indicated that the effect of high-dose DES was not constant across all patients. Specifically, the positive coefficient for treatment-by-age implies that older patients experienced attenuated survival benefits or possible harm compared to younger individuals. Conversely, the negative coefficient for treatment-by-log tumor size indicated that patients with larger tumors derived greater relative benefit from high-dose DES. These interaction patterns were consistent with biological plausibility and clinical expectations, suggesting that the treatment response is conditional on baseline risk factors.

\subsection{Individualized Treatment Effects and Uncertainty Quantification}

To quantify treatment heterogeneity across clinically interpretable subgroups, we estimated patient-specific time ratios from the fitted Weibull AFT model by computing the exponentiated linear predictor differences between the two treatment arms. These individualized treatment effects were evaluated across age and tumor size gradients, with associated 95\% confidence intervals constructed using both the delta method and nonparametric bootstrap procedures.

\begin{figure}[htbp]
    \centering
    \begin{subfigure}[b]{0.7\textwidth}
        \includegraphics[width=\linewidth]{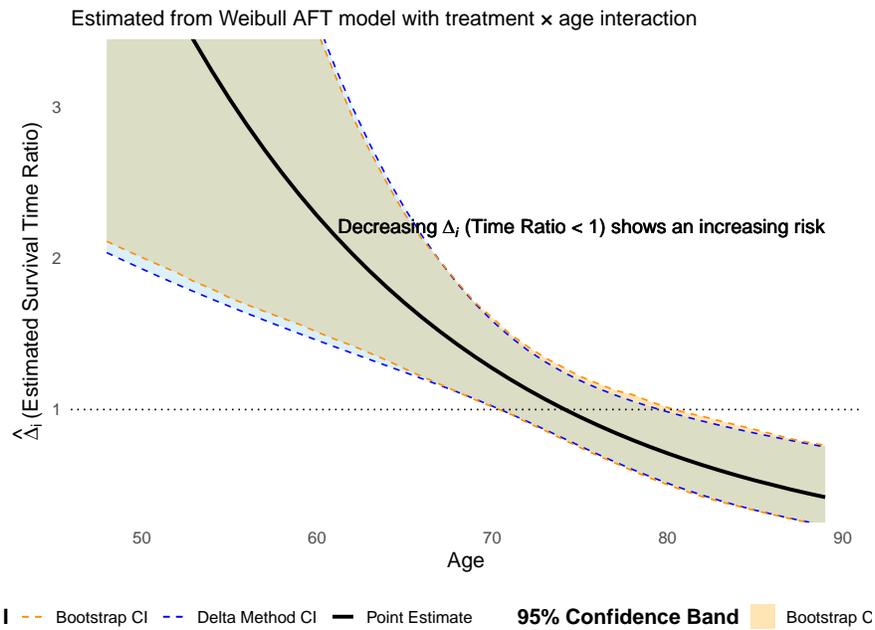}
        \caption{Individualized treatment effects by age. Time ratios above 1 indicate benefit under high-dose DES.}
        \label{fig:age_plot}
    \end{subfigure}
    
    \vspace{1em}
    
    \begin{subfigure}[b]{0.7\textwidth}
        \includegraphics[width=\linewidth]{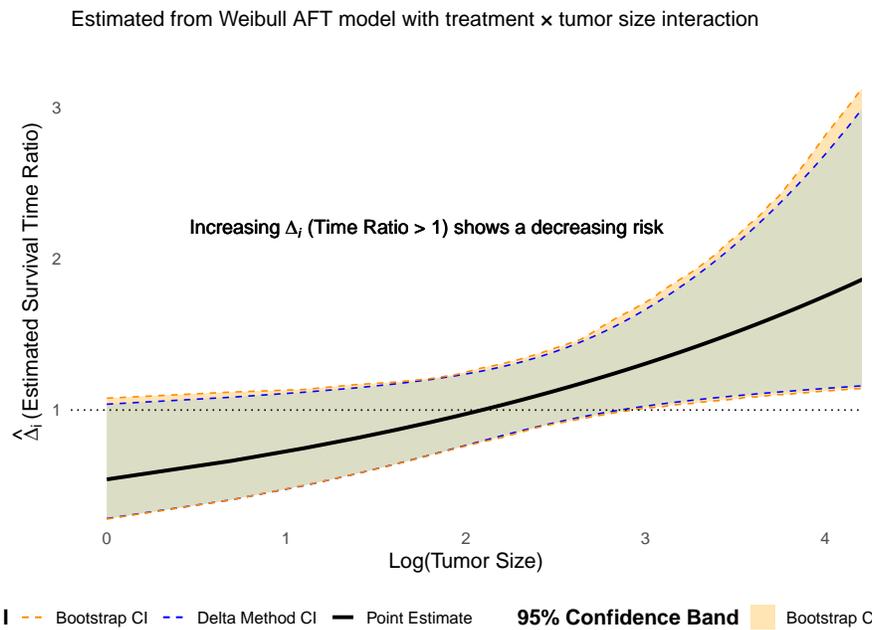}
        \caption{Individualized treatment effects by log tumor size. Time ratios above 1 indicate benefit under high-dose DES.}
        \label{fig:tumor_plot}
    \end{subfigure}

    \caption{Individualized Treatment Effect (ITE) Plots}
    \label{fig:ite_plots}
\end{figure}

Table~\ref{tab:ite-age} and Figure~\ref{fig:age_plot} report treatment effect estimates as a function of age. Among younger patients (e.g., age $\le$ 50 years), the estimated time ratio was 4.09, indicating that the expected survival under high-dose DES was more than four times that under low-dose DES. The 95\% confidence 
The intervals obtained from the delta method ([1.93, 8.67]) and bootstrap resampling ([2.01, 8.60]) both excluded the null value, providing strong evidence of benefit. As age increased, the estimated treatment benefit declined: at age 70 years, the time ratio was 1.27, and for patients aged 80 years and above, the time ratio dropped below 1.0, suggesting possible harm from high-dose DES in these individuals. The agreement between delta-based and bootstrap-based intervals across the age spectrum indicates stable inferences, irrespective of the estimation approach.

\begin{table}[h!]
\centering
\caption{Individualized treatment effects (ITEs) for age with 95\% confidence intervals from delta method and bootstrap.}
\label{tab:ite-age}
\begin{tabular}{cccc}
\toprule
\textbf{Age} & \textbf{exp($\hat{\Delta}_i$)} & \textbf{95\% CI (Bootstrap)} & \textbf{95\% CI (Delta Method)} \\
\midrule
50 & 4.091 & [2.007, 8.601] & [1.931, 8.67] \\
52 & 3.641 & [1.906, 7.195] & [1.828, 7.25] \\
54 & 3.240 & [1.798, 6.017] & [1.73, 6.066] \\
56 & 2.883 & [1.693, 5.007] & [1.636, 5.079] \\
58 & 2.565 & [1.606, 4.208] & [1.546, 4.257] \\
60 & 2.283 & [1.513, 3.506] & [1.458, 3.573] \\
62 & 2.031 & [1.423, 2.941] & [1.373, 3.005] \\
64 & 1.808 & [1.325, 2.496] & [1.289, 2.535] \\
66 & 1.608 & [1.218, 2.126] & [1.204, 2.149] \\
68 & 1.431 & [1.117, 1.843] & [1.116, 1.835] \\
70 & 1.274 & [1.018, 1.608] & [1.023, 1.586] \\
72 & 1.133 & [0.913, 1.418] & [0.921, 1.394] \\
74 & 1.008 & [0.805, 1.282] & [0.813, 1.25] \\
76 & 0.897 & [0.696, 1.173] & [0.706, 1.141] \\
78 & 0.799 & [0.595, 1.099] & [0.604, 1.056] \\
80 & 0.711 & [0.503, 1.013] & [0.513, 0.985] \\
82 & 0.632 & [0.423, 0.955] & [0.433, 0.923] \\
84 & 0.563 & [0.362, 0.896] & [0.364, 0.869] \\
\bottomrule
\end{tabular}
\end{table}

\begin{table}[h!]
\centering
\begin{tabular}{cccc}
\toprule
\textbf{Log Tumor Size} & \textbf{exp($\hat{\Delta}_i$)} & \textbf{95\% CI (Bootstrap)} & \textbf{95\% CI (Delta Method)} \\
\midrule
1.099 & 0.748 & [0.502, 1.136] & [0.500, 1.118] \\
1.386 & 0.813 & [0.577, 1.168] & [0.577, 1.146] \\
1.609 & 0.869 & [0.642, 1.185] & [0.643, 1.173] \\
1.792 & 0.916 & [0.698, 1.206] & [0.700, 1.200] \\
2.303 & 1.065 & [0.853, 1.337] & [0.862, 1.316] \\
3.219 & 1.394 & [1.042, 1.886] & [1.059, 1.835] \\
3.332 & 1.441 & [1.054, 1.978] & [1.074, 1.934] \\
3.555 & 1.539 & [1.085, 2.206] & [1.100, 2.153] \\
3.850 & 1.678 & [1.112, 2.576] & [1.129, 2.494] \\
4.143 & 1.829 & [1.139, 3.031] & [1.155, 2.897] \\
4.248 & 1.887 & [1.144, 3.199] & [1.164, 3.059] \\
\bottomrule
\end{tabular}
\caption{Individualized treatment effects (ITEs) for log tumor size with 95\% confidence intervals from delta method and bootstrap.}
\label{tab:ite-tumor}
\end{table}

Table~\ref{tab:ite-tumor} and Figure~\ref{fig:tumor_plot} show analogous estimates for treatment effect as a function of log tumor size. In patients with small tumors (log size $\approx 1.1$, or $\sim$3 cm\textsuperscript{2}), the estimated time ratio was 0.75, with confidence intervals including the null value, suggesting no significant treatment benefits. As the tumor size increased, the treatment effect became more favorable. At log tumor size 3.3 (approximately 27 cm\textsuperscript{2}), the time ratio exceeded 1.4, and for patients with log tumor size 4.25 (approximately 70 cm\textsuperscript{2}), the time ratio was 1.89 with delta and bootstrap intervals of [1.16, 3.06] and [1.14, 3.20], respectively.

Across both stratifications, the similarity between the inference methods supports the robustness of the estimated ITEs. These results demonstrate clinically interpretable patterns of effect modification and highlight the limitations of relying solely on marginal or average treatment effects.

\subsection{Model-Based Interpretation and Clinical Implications}

Subgroup-specific estimates identified patients who were most likely to benefit from high-dose DES. Specifically, patients under age 65 with tumor areas greater than 25 cm\textsuperscript{2} had estimated time ratios consistently exceeding 2.0, with 95\% confidence intervals that excluded 1.0. This subgroup can be identified through baseline characteristics and represents candidates for intensified therapy. In contrast, patients over 75 years of age with small tumor burdens (less than 10 cm\textsuperscript{2}) showed time ratios below 1.0, indicating potential overtreatment with high-dose DES.

The ability to identify such subgroups through parametric modeling with interaction terms represents a meaningful extension of the conventional survival analysis. Rather than relying on population-averaged effects, individualized time ratios derived from flexible parametric models allow tailored risk-adaptive treatment strategies. These estimates provide an interpretable framework for incorporating patient heterogeneity into clinical decision-making for advanced prostate cancer.

\section{Discussion}

The modeling strategy adopted in this analysis represents a targeted effort to quantify the heterogeneity in survival outcomes under different doses of diethylstilbestrol (DES) among men with advanced prostate cancer. The Weibull accelerated failure time (AFT) model, specified with targeted treatment–covariate interactions, enabled the estimation of individualized treatment effects (ITEs) that capture patient-specific survival contrasts, offering a more granular alternative to conventional average treatment effect approaches. Rather than treating heterogeneity as a nuisance, this framework formally incorporated clinical effect modifiers (age and tumor size) into the survival model, allowing for clinically interpretable and statistically valid time ratio estimates tailored to individual covariate profiles.

Substantial variation in treatment efficacy was observed. Among men aged 50–60 years, high-dose DES was associated with substantial survival gains, with estimated time ratios exceeding 4.0 and non-overlapping confidence intervals via both delta method and bootstrap estimation. In contrast, the survival benefit diminished markedly with advancing age, with estimates crossing or falling below unity in those aged $\ge$ 75 years. A similar pattern emerged for tumor burden: patients with larger tumors (e.g., tumor area $>$ 27 cm\textsuperscript{2}) demonstrated a clear benefit from high-dose treatment, while those with small tumors showed no discernible gain. These findings support the hypothesis that a uniform dosing strategy fails to capture clinically important subpopulation effects and further suggest the presence of negative treatment effects in low-risk groups.

The results reinforce and extend previous studies on covariate-treatment interactions in survival settings. While randomized trials often report hazard ratios as a summary measure, such estimates assume proportional hazards and may obscure time-dependent or subgroup-specific effects \citep{hernan2010hazards, aalen2015survival}. In contrast, the AFT modeling framework provides multiplicative effects on the survival time scale and accommodates flexible interaction. Previous findings on DES efficacy in prostate cancer trials have reported conflicting conclusions \citep{bosland2005chemoprevention}, but rarely accounted for variations across patient strata. The present analysis shows that such variation is not only present but also consequential.

As a methodological contribution, this study also demonstrates a practical implementation of individualized causal effect estimation under a potential outcomes framework using parametric survival models. Both the delta method and nonparametric bootstrap were used for ITE uncertainty quantification. Their near-identical results across subgroups emphasize the reliability of the estimation procedure. In this sense, the modeling pipeline serves as a proof-of-concept for the application of individualized survival modeling in oncology trials, particularly when the objective is to identify response heterogeneity rather than merely testing for overall treatment efficacy.

This study has several important limitations. The analysis relied on a moderately sized historical randomized dataset, with possible imbalances in sparsely represented covariate combinations. To ensure model stability, the interaction structure was deliberately restricted to clinically plausible and statistically supported terms, and other covariates were excluded based on convergence diagnostics and model selection criteria. While the Weibull AFT model provided a favorable fit based on the AIC, BIC, and graphical diagnostics, other survival models (e.g., log-normal and flexible parametric) may offer improved robustness in alternative settings. External validation in more contemporary cohorts is required to assess generalizability, particularly under modern treatment protocols.

Nonetheless, the results illustrate the potential value of individualized time-ratio estimation for guiding treatment selection. When covariate-treatment interactions are well-characterized and interpretable, model-based ITEs provide an evidence-driven path toward risk-adaptive decision-making. Rather than forcing a binary treatment assignment across a heterogeneous population, precision modeling offers the opportunity to tailor therapeutic intensity based on the expected survival gains. Such approaches may be especially relevant in oncology, where treatment toxicity profiles differ across age and disease stage, and where one-size-fits-all strategies often underperform in real-world settings.

\section*{Conflict of Interest Statement}

The authors declare that the research was conducted in the absence of any commercial or financial relationships that could be construed as potential conflicts of interest.

\section*{Data Source}
Byar, D.P. and Green, S.B. (1980). The choice of treatment for cancer patients based on covariate information: applications to prostate cancer. Bulletin du Cancer 67: 477-490

\newpage
\bibliographystyle{apalike}
 
\bibliography{test}

\newpage
\section*{APPENDIX}

\begin{table}[h!]
\centering
\begin{tabular}{lll}
\toprule
\textbf{Variable} & \textbf{Description} & \textbf{Wald test} \\
\midrule
\textbf{Continuous Covariates} & & \\
age & Age at diagnosis & 0 \\
wt & Weight (in kg) & 0.008 \\
sbp & Systolic Blood Pressure & 0.669 \\
dbp & Diastolic blood pressure & 0.43 \\
sz & Size of primary tumor (cm\textsuperscript{2}) & 0 \\
ap & Serum Acid Phosphatase & 0.581 \\
hg & Hemoglobin (in g/100 ml) & 0 \\
sg & Combined index of tumour stage \& histologic grade & 0 \\
\midrule
\textbf{Categorical Covariates} & & \\
pf & Performance status & 0 \\
hx & History of cardiovascular disease & 0 \\
bm & Presence of bone metastases & 0 \\
stage & Stage 4 vs. stage 3 & 0.002 \\
ekg & Abnormal electrocardiogram & 0.001 \\
\bottomrule
\end{tabular}
\caption{Univariate analysis of clinical covariates based on Wald tests from the Cox model. Covariates with low p-values were prioritized for model inclusion.}
\label{tab:Table 2}
\end{table}

\newpage
\begin{table}[h!]
\centering
\begin{tabular}{lcc|c}
\toprule
 & Death & Alive & Total \\
\midrule
Low Dose (Control Group) & 180 (37.89\%) & 57 (12\%) & 237 (49.89\%) \\
High Dose (Treatment Group) & 158 (33.26\%) & 80 (16.84\%) & 238 (50.11\%)\\
\midrule
Total  &  338 (71.58\%) & 137 (28.42\%) & 475 (100\%) \\
\bottomrule
\end{tabular}
\caption{Low Dose vs. High Dose Frequencies}
\label{tab:treat_diag}
\end{table}

\begin{table}[ht]
\centering
\caption{Model comparison based on Akaike Information Criterion (AIC) and Bayesian Information Criterion (BIC).}
\begin{tabular}{lcc}
\toprule
\textbf{Model} & \textbf{AIC} & \textbf{BIC} \\
\midrule
Log-normal     & 3300.081 & 3354.204 \\
Log-logistic   & 3276.541 & 3330.664 \\
Weibull        & 3260.219 & 3314.342 \\
\bottomrule
\end{tabular}
\label{tab:model_comp}
\end{table}

\begin{table}[h!]
\centering
\caption{Model comparison (Weibull AFT) using Akaike (AIC) and Bayesian (BIC) Information Criteria. Lower values indicate better fit.}
\begin{tabular}{lrr}
\toprule
\textbf{Model} & \textbf{AIC} & \textbf{BIC} \\
\midrule
Model 1  & 3273.508 & 3319.304 \\
Model 2A & $\mathbf{3260.219}$ & $\mathbf{3314.342}$ \\
Model 2B & 3269.744 & 3348.847 \\
\bottomrule
\end{tabular}
\label{tab:model_comp_poster}
\end{table}

\end{document}